\documentclass[mathleft
]{an}
\usepackage{graphicx}
\usepackage{times}
\begin{document}

\Pagespan{789}{}
\Yearpublication{2006}%
\Yearsubmission{2005}%
\Month{11}%
\Volume{999}%
\Issue{88}%

\title{Asteroseismic determination of the physical characteristics of the planetary system host HR\,8799 ($\lambda$ Bootis nature and age)}

\author{A. Moya\inst{1}\fnmsep\thanks{Corresponding author:
  \email{amoya@cab.inta-csic.es}\newline}, P. J. Amado\inst{2}, D. Barrado\inst{3,1}, A. Garc\'{\i}a Hern\'andez\inst{2}, M. Aberasturi\inst{1}, B. Montesinos\inst{1}, F. Aceituno\inst{2}}

\titlerunning{Asteroseismic determination of the physical
  characteristics of the planetary system host HR\,8799 ($\lambda$
  Bootis nature and age)} \authorrunning{A. Moya et al.}  \institute{
  LAEX-CAB, Departamento de Astrof\'{\i}sica, Centro de
  Astrobiolog\'ia (INTA-CSIC), PO BOX 78, 28691 Villanueva de la
  Ca\~nada, Madrid, Spain \and Instituto de Astrof\'isica de
  Andaluc\'ia, IAA - CSIC, Granada, Spain E-18008 \and Calar Alto
  Observatory, German-Spanish Astronomical Center, C/ Jes\'us
  Durb\'an Rem\'on, 2-2, 04004, Almer\'ia, Spain } 


\abstract{HR\,8799 is a $\lambda$ Bootis, $\gamma$ Doradus star
  hosting a planetary system and a debris disk with two rings. This
  makes this system a very interesting target for asteroseismic
  studies. In particular, this work is devoted to the determination of
  the internal metallicity of this star, linked with its $\lambda$
  Bootis nature, and its age, taking the advantage of its $\gamma$
  Doradus-type pulsations. To do so we have used the equilibrium code
  CESAM and the non-adiabatic pulsational code GraCo. We have applied
  the Frequency Ratio Method and the Time Dependent Convection theory
  to estimate the mode identification, the Brunt-Va\"is\"al\"a
  frequency integral and the mode instability, making a selection of
  the possible models fulfilling all observational constraints. Using
  the position of the star in the HR diagram, the solar metallicity
  models is discarded. This result contradicts one of the main
  assumptions of the most accepted hypothesis explaining the $\lambda$
  Bootis nature, the accretion/diffusion of gas from a star with solar
  metallicity. Therefore, in sight of these new results, a revision of
  this hypothesis is needed. The inclusion of accurate internal
  chemical mixing is necessary. The use of the asteroseismological
  constraints provides a very accurate determination of the physical
  characteristics of HR\,8799: an age in the ranges [1123, 1625] and
  [26, 430] Myr, and a mass in the ranges [1.32, 1.33] and [1.44,
    1.45] $M_{\odot}$, respectively, depending on the visual angle
  $i$. The determination of this angle and more accurate multicolor
  photometric observations can definitively fix the mass, metallicity
  and age of this star. In particular, an accurate age estimation is
  needed for a correct understanding of the planetary system. In our
  study we have found that the age widely used for modelling the system
  is unlike.}

\maketitle

\section{Introduction}


Accurate ages stars hosting extrasolar planets are useful for, e.g.,
obtaining theoretical constraints on tidal interactions between Hot
Jupiters and their hosts, dynamical modelling of exoplanetary systems,
predicting exoplanetary luminosities, etc, since they constrain the
age of the planets. The eventual aim of refining the age determination
of the stars is to obtain more information on planetary system
formation to discern between the different scenarios. Several methods
can be applied to determine ages of stars, like the use of
chromospheric activity proxies, lithium depletion, evolutionary track
fitting, etc (Barrado y Navascu\'es 1998; Barrado y Navascu\'es et
al. 1999). Nonetheless, for pulsating stars, asteroseismology is the
best and most accurate method we know that provides mass, radii and
ages with an accuracy order of magnitudes larger than obtained with
other methods in the literature (Stello et al. 2009).

The discovery by Marois et al. (2008), of the first planetary system by
direct imaging around HR\,8799 has made the age, mass and metallicity
determination of this star a very important task. That was the
starting point of a set of approximately ten studies about this system
during 2009, none of them from an asteroseismic point of view, even
considering that there are three detected pulsational frequencies
(Zerbi et al. 1999) that can be used to better understand the host
star.

The main characteristics of this system are:

\begin{itemize}

\item The host star is a A5 Main Sequence star (Gray \& Kaye 1999),
$\gamma$ Doradus pulsator (Zerbi et al. 1999), and it has $\lambda$
Bootis-type surface chemical peculiarities (Sadakane 2006).

\item There are three sub-stellar objects orbiting the host star. They
have been observed using direct imaging (Marois et al. 2008).

\item The system has a debris disk with two rings, one inner and another
further out the three orbiting objects (Chen et al. 2009).

\item From a dynamical point of view, the stability of the system is a
challenge, regarding the mass given for the objects and star, and the
age assumed for the system (Reidemeister et al. 2009).

\end{itemize}

These four characteristics make the planetary system HR\,8799 a very
interesting one, with many challenges to understand its structure,
formation and evolution. In particular, the age of the star (and
consequently of the system) is one of the key quantities for all the
studies done around HR\,8799, since it is directly related with the
mass of the orbiting objects, and it imposes a goal for the dynamical
stability of the system.

In two recent works (Moya et al., 2010a, b), a comprehensive modelling
of HR8799 using asteroseismology has been done. The main conclusions
are the following:

\begin{itemize}
\item The use of asteroseismological observational data imposes large
  constraints to the acceptable models of stellar internal structure.

\item The actual uncertainty in the knowledge of the age of HR\,8799
  is larger than shown in the literature. Only less than 20\% of the
  accepted model using all the observational constraints have an age
  in the range estimated by other works.

\item The internal subsolar metallicity obtained for the star provides
some keys to understand the $\lambda$ Bootis nature.
\end{itemize}

These works have shown the enormous potential of the use of
asteroseismology to understand this system and others. Nevertheless,
these studies cannot be conclusive due to the poor asteroseismic
observational available data. There is not very accurate multicolour
photometric observations. Therefore, additional observations are
imperative to provide better asteroseismological constraints.

\section{Observational data}

The A5 V star HR\,8799 has been extensively studied in the last
years. The first studies of this star were done in the context of
asteroseismology. Schuster \& Nissen (1986) firstly reported HR\,8799
as a possible SX Phoenicis type. Zerbi et al. (1999) observed this
star in a multisite multicolour photometric campaign, with Str\"omgren
filters, and found three independent pulsational frequencies
($f_1=1.9791\;\rm{c}\,\rm{d}^{-1}$ , $f_2=1.7268\;\rm{c}\,\rm{d}^{-1}$
, and $f_3=1.6498\;\rm{c}\,\rm{d}^{-1}$ , units are cycles per day),
making it one of the 12 first $\gamma$ Doradus pulsators known (Kaye
et al 1999). This pulsating stellar group is composed of Main Sequence
(MS) stars in the lower part of the classical instability strip. Their
pulsation modes have periods in the range [0.5,3] days, that is, they
are asymptotic g-mode pulsators. Gray \& Kaye (1999) obtained an
optical spectrum of HR\,8799, and assigned an spectral type of kA5 hF0
mA5 V $\lambda$ Bootis, reporting an atmospheric metallicity of
[M/H]=$-0.47$ and accurate values of the stellar luminosity,
$T_{\rm{eff}}$, and $\log\,g$ (see Table 1) . They also noted that
HR\,8799 may be also a Vega-type star, characterized by a far IR
excess due to a debris disk. Sadakane (2006) developed a deep study of
the metal abundances of this star, confirming its $\lambda$ Bootis
nature (with surface chemical peculiarities, Paunzen 2003). HR\,8799
is one of the three $\lambda$ Bootis stars with $\gamma$ Doradus
pulsations (Rodr\'iguez et al. 2006a, b)

The use of different bolometric corrections changes significantly the
value of the luminosity of the star. Using the Virtual observatory
tool VOSA (Bayo et al. 2008), models with realistic metallicities best
fitting the observations provide a luminosity only 0.1 $L/L_{\odot}$
larger than that given in Table 1 and, therefore, within the
errors. On the other hand, the use of different parallaxes in the
literature changes the value of the absolute magnitude of this star by
less than 0.1 $L/L_{\odot}$ (see Moya et al. 2010a for details).

\begin{table}
 \centering
  \caption{Physical characteristics of HR\,8799}
  \begin{tabular}{rrr}
 \hline
$T_{\rm{eff}}$ (K) & 7430$\pm$75 & \\
$\log g$ ($\rm{cm}\,\rm{s}^{-2}$) & 4.35$\pm$0.05 & Gray \& Kaye 1999\\
$\rm{L} (L_\odot)$ & 4.92$\pm$0.41 & \\
\hline
$\rm{v}\sin i$ (km $\rm{s}^{-1}$) & 37.5$\pm$2 & Kaye \& Strassmeier 1998\\
\hline
$\pi$ (mas) & 25.38$\pm$0.85 & van Leeuwen 2007\\
\hline
\end{tabular}
\end{table}



\section{Modelling of the observational data}

We have developed a grid of equilibrium models obtained with the CESAM
code (Morel \& Lebreton 2008). We vary the mass (in the range [1.25,
  2.10] $M_\odot$ with steps of 0.01 $M_\odot$), the metallicity (with
values [M/H]=0.08, $-$0.12, $-$0.32, and $-$0.52), the Mixing-Length
parameter MLT (values 0.5, 1, and 1.5), and the overshooting (values
0.1, 0.2, and 0.3). The internal metallicity has been regarded as a
free parameter due to the $\lambda$ Bootis nature of the star, which
hides its internal abundances. The mass, estimated to be $1.47\pm0.3$
$M_\odot$ by Gray \& Kaye 1999, has been also regarded as a free parameter
since it has not been directly determined.

The pulsational analysis has been developed using the non-adiabatic
pulsational code GraCo (Moya et al. 2004; Moya \& Garrido 2008). Both
codes have been tested in the work of the ESTA activity (Evolution and
Asteroseismic Tools Activities, Lebreton et al. 2008; Moya et
al. 2008). Using these tools, we performed a theoretical study of
HR\,8799 in an attempt at constraining physical and theoretical
parameters. In this work we will follow the same scheme used for the
study of the $\delta$ Scuti pulsators RV Arietis, 29 Cygnis and 9
Aurigae (Casas et al. 2006; Casas et al. 2009; Moya et
al. 2006). Taking advantage of the $\gamma$ Doradus pulsations, we
have used the Frequency Ratio Method (Moya et al. 2005) to estimate
the possible mode identification of the observed modes, and the Time
Dependent Convection (Gricahc\'ene et al. 2005) for the analysis of
the mode instability and the possible spherical order comparing with
observed multicolour photometry. All these techniques provide
additional constraints to the position of the star in the HR diagram,
reducing the possible physical characteristics of the models
fulfilling all the observations.

\section{On the $\lambda$ Bootis nature of HR\,8799}

The first selection of models was done imposing only the spectroscopic
constraints displayed in Table 1. This selection shows that there are
no models with solar metallicity fulfilling observations, since the
stellar luminosity derived is smaller than any of the possible
luminosities of models with solar metallicity. This contradicts the
main assumption of the theories explaining the $\lambda$ Bootis
nature, i.e. the accretion/diffusion scenario where these stars have
solar metallicity, whereas the observed abundances are due to surface
phenomena (Turcotte \& Charbonneau 1993).

The result of our work, discarding solar metallicity as the internal
metallicity of the star, together with the fact that some $\lambda$
Bootis stars have debris disks not connected with the star (Chen et
al. 2009), makes this accretion/diffusion scenario unlikely, but not
negligible (Su et al. 2009).  Therefore, the study of internal
chemical mixing processes seems to be the key to explain the $\lambda$
Bootis nature, at least for HR\,8799, as the solar abundances for C,
N, O and S observed on its surface have still to be explained.

\section{On the mass and age of HR\,8799}

There are several age determinations of the age of HR\,8799 in the
literature (see Moya et al. 2010b for a complete review), most of them
estimating its age in a range around [30,160] Myr. The most roboust
determination found is the kinematic age determination (UVW). But the
authors warned that this method is not always reliable.

\begin{figure}
\includegraphics[width=80mm]{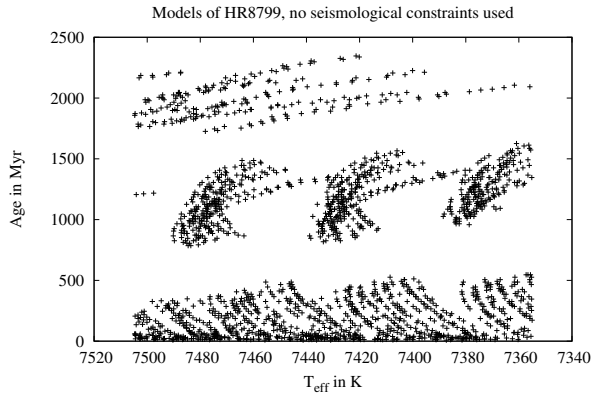}
 \caption{$T_{\rm{eff}}$ - Age diagram for the models of our grid
   fulfilling the spectroscopic observations, i.e., $T_{eff}$,
     $\log g$ and luminosity displayed in Table 2. The age range
     obtained is [10, 2337] Myr.}
\end{figure}

The main complementary argument to that statistical determination is
the position of the star in the HR diagram. In Moya et al. 2010b we
have demostrated that this procedure does not provide accurate age
estimations due to the $\lambda$ Bootis nature of HR\,8799
(Fig. 1). This result was previusly pointed out by Song et al. (2001),
how stimated the age of HR\,8799 in the range [50,1128] Myr. In
Section 4 we saw that this nature hides the real internal metallicity
of the star. The main consequence of this result is that the models
fulfilling observations are in a range of ages [10,2337] Myr, a much
broader range that one estimated by other authors of [30,160] Myr
(Marois et al. 2008). Only a small amount (18.1$\%$) of models in our
representative grid have ages in the range mainly claimed in the
literature.

\begin{figure}
\includegraphics[width=80mm]{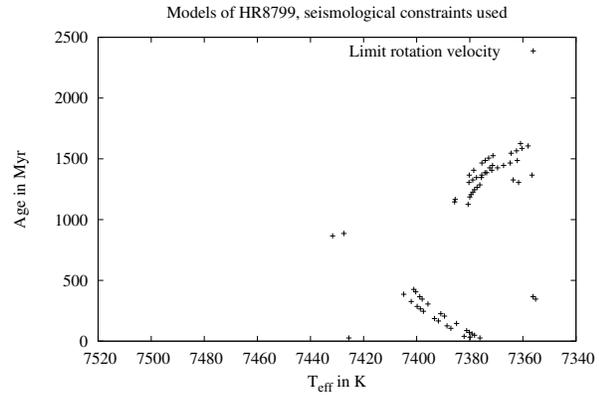}
 \caption{$T_{\rm{eff}}$ - Age diagram for the models of our grid
   fulfilling the spectroscopic observations plus all the
   asteroseismological constraints (FRM + multicolour photometry +
   instability analysis, see Moya et al. 2010a, b) for a rotation
   velocity $V_{\rm{rot}}\approx 60 \rm{km}\,\rm{s}^{-1}$. The age
   ranges obtained are $[1123, 1625]$, $[26, 430]$ Myr}
\end{figure}

Therefore we need aditional constraints for an accurate estimation of
the age and mass of this star. We have explained in Section 3 that one
of the techniques used for this study is the Frequency Ratio
Method. This method has a range of applicability deppending on the
stellar rotation velocity (Su\'arez et al. 2005; Moya et
al. 2010a). For inclination angles around $i=36^\circ$ (corresponding
to a stellar rotation velocity of $V_{\rm{rot}}\approx 60
\rm{km}\,\rm{s}^{-1}$ (Kaye \& Strassmeier 1998), the limit for the
FRM to be accurately applied), the models fulfilling all the
observational constraints have masses in two separate ranges of
M=[1.32, 1.33], [1.44, 1.45] $M_{\odot}$. The age of the system is
constrained in two separate ranges: [1123, 1625] Myr and [26, 430] Myr
respectively (Fig. 2). A percentage of 16.7$\%$ of the models are in
the range given in the literature. A consequence of this result is
that, in the case of the youngest age range, the predicted masses of
the observed planets are [5,14] $M_{\rm{Jup}}$ for the most luminous
planets, and [3,13] $M_{\rm{Jup}}$ for the less luminous one. The
oldest age range, the most probable from the point of view of the
present study, predicts masses for the three objects in the brown
dwarfs domain (see Fig. 4 of Marois et al. 2008).

\begin{table*}
 \centering
  \caption{Acceptable models depending on the physical constraints. In
    the complete procedure case, the first (second) mass range is
    linked with the first (second) age range shown.}
  \begin{tabular}{cccc}
  \hline Constraint & Mass & Age & $\%$ of models with $[30, 160]$ \\
 &  in $M_{\odot}$ & in Myr & in Myr \\
\hline 
HR position & [1.25,1.27],[1.32,1.35],[1.40,1.48] & [10, 2337] & 18.1\\ 
Complete procedure & [1.32, 1.33], [1.44, 1.45] & $[1123, 1625]$, $[26, 430]$ & 16.7\\ 
\hline
\end{tabular}
\end{table*}

The lack of an accurate determination of the inclination angle is the
main source of uncertainty of the present study.  This angle has not
been unambiguously obtained up to now, and its value would say whether
the results of this study are actually applicable, and then, it can
provide a very accurate determination of the age and mass of the
star.

\section{Conclusions}

The accurate determination of the mass, age and internal metallicity
of a star hosting a planet or a planetary system is a necessary step
to understand the formation and evolution of that planetary system. If
the hosting star pulsates, the use of asteroseismology can provide
these physical quantities with an accuracy hardly obtained with other
techniques.

The case of HR\,8799 is a excellent example of this benefit provided
by asteroseismology. In two recent works (Moya et al. 2010a, b), the
first comprehensive asteroseismologic study of the planetary system
host HR\,8799, a $\lambda$ Bootis star presenting $\gamma$ Doradus
pulsations has been carried out. This asteroseismic work is specially
important for the determination of the internal abundances of this
kind of stars, a previous step to understand the physical mechanism
responsible for the surface chemical peculiarities of the $\lambda$
Bootis group. On the other hand, an analysis of the age determination
of the planetary system HR\,8799 has also been done. The results found
in the literature are not conclusive, and the only valid argument to
estimate the age of the star is that using its radial velocity and
proper motion, but it is an estatistical argument needed of additional
estimations.

The selection of models fulfilling the spectroscopic observations
shows that there are no selected models with solar metallicity. This
has an impact in the main assumption of the theory widely accepted to
explain the $\lambda$ Bootis nature, i.e. that these stars have solar
metallicity, whereas the observed abundances are due to surface
phenomena.

On the other hand, the main complementary argument to the kinematic
age determination is the position of the star in the HR diagram. We
have shown that, due to the $\lambda$ Bootis nature of HR\,8799, that
hides the internal metallicity of the star, this HR diagram position
provides a range of ages [10,2337] Myr, a much wider than that
estimated by other authors of [30,160] Myr (Marois et al. 2008). Only
a small amount (18.1$\%$) of models in our grid have ages in the range
claimed in the literature.

For inclination angles around $i=36^\circ$, the models fulfilling all
the observational constraints have masses in two separate ranges of
M=[1.32, 1.33], [1.44, 1.45] $M_{\odot}$, and the age of the system is
constrained in two separate ranges: [1123, 1625] Myr and [26, 430] Myr
respectively. A percentage of 16.7$\%$ of the models are in the range
given in the literature. This determination has an impact in the
determination of the mass of the objects observed orbiting around
HR\,8799.

A consequence of this study is the need for a precise determination of
the inclination angle $i$, of the multicolour photometric amplitudes
and phases of $f_2$, and some information of $m$ values through
time-series if high resolution spectroscopy. These determinations
would help to carry out a definitive selection of the models. In any
case, the range of ages assigned to this star in the literature is
unlikely to be the correct one. Only a stellar luminosity larger than
that reported would allow young models with solar metallicity to
fulfill all the observational constraints.

\acknowledgements {PJA acknowledges financial support from a ``Ramon y
  Cajal'' contract of the Spanish Ministry of Education and
  Science. This research has been funded by Spanish grants
  ESP2007-65475-57-C02-02 , CSD2006-00070, ESP2007-65480-C02-01,
  AYA2009-08481-E and CAM/PRICIT-S2009ESP-1496.}


\end{document}